\documentstyle[NATO,numreferences,epsfig]{crckapb}
\begin{opening}
\title{Near-Field Microwave Microscopy of Materials Properties }
\author{Steven M. Anlage}
\author{D. E. Steinhauer}
\author{B. J. Feenstra}
\author{C. P. Vlahacos}
\author{F.\ C. Wellstood}

\institute{Center for Superconductivity Research, and 
Materials Research Science and
Engineering Center, 
Department of Physics, 
University of Maryland, 
\mbox{College Park, Maryland 20742-4111 USA}}

\end{opening}

\begin{document}

\begin{abstract}
Near-field microwave microscopy has created the opportunity for a new class
of electrodynamics experiments of materials. Freed from the constraints of
traditional microwave optics, experiments can be carried out at high spatial
resolution over a broad frequency range. In addition, the measurements can
be done quantitatively so that images of microwave materials properties can
be created. We review the five major types of near-field microwave
microscopes and discuss our own form of microscopy in detail. Quantitative
images of microwave sheet resistance, dielectric constant, and dielectric
tunability are presented and discussed. Future prospects for near-field
measurements of microwave electrodynamic properties are also presented.
\end{abstract}

\section{Introduction}

Measurement of the electromagnetic response of materials at microwave
frequencies is important for both fundamental and practical reasons. The
complex conductivity of metals and superconductors gives insights into the
physics of the quasiparticle excitations and collective charge properties of
the material. Interesting collective behavior which can be investigated
through the complex conductivity includes superconductivity, spin and charge
density waves, Josephson plasmons, etc. The dielectric properties of
materials give insights into the polarization dynamics of insulators and
ferroelectrics. A class of interesting insulating materials, the parent
compounds of cuprate superconductors, combine both dielectric and
antiferromagnetic behavior because of strong correlations between the charge
carriers. Ferromagnetic materials interact with microwaves in several
unusual ways, including ferromagnetic resonance and anti-resonance, and
spin-wave resonance. In addition, a series of discoveries over the past
several decades have found many materials which display coexistence of
superconductivity with either antiferromagnetism, or modified forms of
ferromagnetism. This wide range of interesting electromagnetic behavior of
contemporary materials requires that experimentalists working in this field
master many diverse measurement techniques and have a broad understanding of
condensed matter physics.

On the practical side, electromagnetic measurements are essential for
creating new device technologies and optimizing existing devices and
processes. For example, nonlinearities in superconducting materials limit
their applications at microwave frequencies to relatively low power uses.
Although intrinsic nonlinearities of superconductors will eventually limit
their utility, most practical nonlinearities are caused by extrinsic defects
in the material or device. Careful characterization of these materials on
the length scale of the extrinsic inhomogeneities is required to tackle this
problem. Another practical issue is the optimization of materials for
frequency-agile applications. Here the issue is measurement of dielectric or
magnetic properties which can be tuned with an electric or magnetic field,
often in thin-film materials. Yet another important application of
electromagnetic measurements is in diagnostic measurements. For example, in
a semiconductor integrated circuit process, one needs to measure and control
quantities such as sheet resistance, doping profiles, and electromigration
in wires. As the speed of microprocessors continues to climb, the microwave
properties of materials become increasingly important. Finally, if
high-speed integrated circuits fail, it is important to locate the fault
(often an open or short) as accurately as possible, and to correct the
problem quickly to maintain a high yield. Electromagnetic measurements are
thus an important foundation for many emerging technologies.

In this paper we present a new paradigm for electromagnetic measurements of
materials. We first briefly review the traditional methods of microwave
measurements and point out some of their important limitations. We then
present an alternative approach to these measurements through quantitative
near-field microwave microscopy. The remainder of the paper is devoted to
reviewing the progress we have made in this exciting new field of research.
\clearpage
\section{ Traditional Microwave Measurements of Electromagnetic Properties}

The fundamental electrodynamic quantities of greatest interest are the
surface impedance Z$_s$, the conductivity $\sigma $, the dielectric
permittivity $\varepsilon $, and the magnetic permeability $\mu $. A review
of the definitions of conductivity and surface impedance of normal metals
and superconductors is given elsewhere in this book. All of these quantities
are complex and in general are a function of many variables, including
frequency, temperature, magnetic field, electric field, etc.

Traditional microwave measurements of these quantities are typically done on
length scales of the free-space wavelength of the signal, which is about 3
cm at 10 GHz. The earliest microwave measurements on superconductors by
Pippard were done with the sample acting as a quarter-wavelength resonator 
\cite{Pippard}. In this case the electromagnetic properties measured are
actually an average of the properties along the sample, weighted by the form
of the standing-wave resonance pattern. In the case of elemental Pb or Sn
this is not much of an issue, but in the case of complicated multi-element
oxides it can give rise to misleading results.

A refinement of this technique came through the use of cavity perturbation
methods. In this case the sample is immersed in a larger electromagnetic
cavity in a region of uniform electric or magnetic field \cite{Slater,UCLA}.
The properties of the sample can be deduced by comparing the resonant
frequency shift and quality factor change from a well-characterized initial
(unperturbed) state of the cavity. In this case one measures an average of
the electromagnetic properties of the sample, again weighted by the
distribution of fields and currents created in the sample \cite{UCLA}. These
field and current distributions can be quite complicated for most practical
samples \cite{SC,Gough}, becoming simple only for carefully shaped
ellipsoids of revolution.

Other resonant techniques have been developed in recent years, particularly
for examination of thin-film superconducting materials. These include the
parallel-plate resonator technique, in which two congruent films form a
transmission line resonator with a length scale on the order of the
microwave wavelength \cite{Taber,Talanov}. Dielectric resonators are also
sensitive to thin-film surface impedance but are averaged over an area on
the order of the wavelength \cite{Gallop,Wilker93}. Far-field
surface-impedance microscopy techniques use a diffraction-limited beam in a
confocal geometry to locally measure the surface impedance at
millimeter-wave frequencies, but have a spatial resolution limited to a few
millimeters \cite{Hogan,Martens}.

Non-resonant methods can also be employed to measure electrodynamic
properties. For instance, waveguide transmission through thin films has been
successful at determining the absolute value of the penetration depth, but
provides an average of the properties over the film, again on the length
scale of the microwave wavelength.

\subsection{Limitations of Traditional Techniques}

Although traditional electrodynamics measurement techniques have been very
successful, they do suffer from some fundamental limitations. First, they
tend to measure a weighted average over large areas of the sample, on the
scale of millimeters to centimeters. In the case of oxide thin films, it is
often difficult to prepare a material which is homogeneous on such long
length scales. For oxide single crystals, doping inhomogeneities have been
shown to exist on many length scales from the nanometer to the millimeter
range \cite{NRLScience}. In addition, surface morphology and second phases
(such as flux particles) complicate the response of the crystal to
electromagnetic radiation. Hence, traditional electrodynamics measurements
yield only an average or overall picture of the sample properties. This
means that the intrinsic behavior of the material can be masked by the
response of a relatively small fraction of extrinsic or second-phase
material. When new materials are measured, there is often a lingering doubt
whether the response is due to new intrinsic physics or some unforeseen
extrinsic effect.

A second limitation of traditional techniques is the generation of large
screening currents, especially near edges, and in the case of
superconductors, the subsequent admission of rf magnetic flux into the
sample. The rf flux will significantly increase the surface impedance and
can easily dominate the response of the material. Such issues have arisen
many times in the exploration of electromagnetic properties of new
superconducting materials. For instance, early measurements of the
penetration depth in thin films of YBCO were corrupted by vortex entry into
the films \cite{HebardDefects}; harmonic generation in superconducting
crystals can be dominated by edge effects \cite{Hampel}; and the nonlinear
Meissner effect in superconductors can be swamped easily by vortex entry
problems \cite{Hardy,Carr}.

\subsection{A new paradigm for electrodynamics measurements}

To overcome the limitations of traditional electrodynamics measurements, we
have developed a new family of near-field microwave microscopes which permit
local quantitative measurements of surface impedance and conductivity, as
well as dielectric and magnetic properties. Near-field techniques also
relieve us from the ``light-cone constraint'', in which the length scales
that we can probe with electromagnetic radiation are dictated by the
frequency. For instance, this permits broadband electrodynamics measurements
to be performed on very fine length scales. Because the spatial resolution
is determined by geometrical features which we can control, a class of
altogether new electromagnetic experiments can be performed. For example, we
now can measure the local tunability of high-frequency electrical properties
and make new connections between microstructure and its associated physical
properties.

\section{An Overview of Microwave Microscopy Techniques}

Near-field microwave microscopy is an art which has formed gradually over
the years from many different sources. In principle the intellectual founder
of the near-field microscopy effort was Synge in his prescient paper of 1928 
\cite{Synge}. However, most practitioners of the art are not familiar with
this seminal work despite its importance. Many near-field techniques were
developed empirically, some, in fact, developed to solve practical problems
in manufacturing. The earliest effort to perform high resolution
quantitative microwave measurements seems to be from the ferromagnetic
resonance community, led by the work of Frait \cite{Frait}, and later Soohoo 
\cite{Soohoo}. However, not all subsequent developments can be traced back
to these roots.

\begin{figure}[h]
%\vspace{5cm}   % amount of vertical space needed
\par
\begin{center}
\leavevmode
\epsfig{file=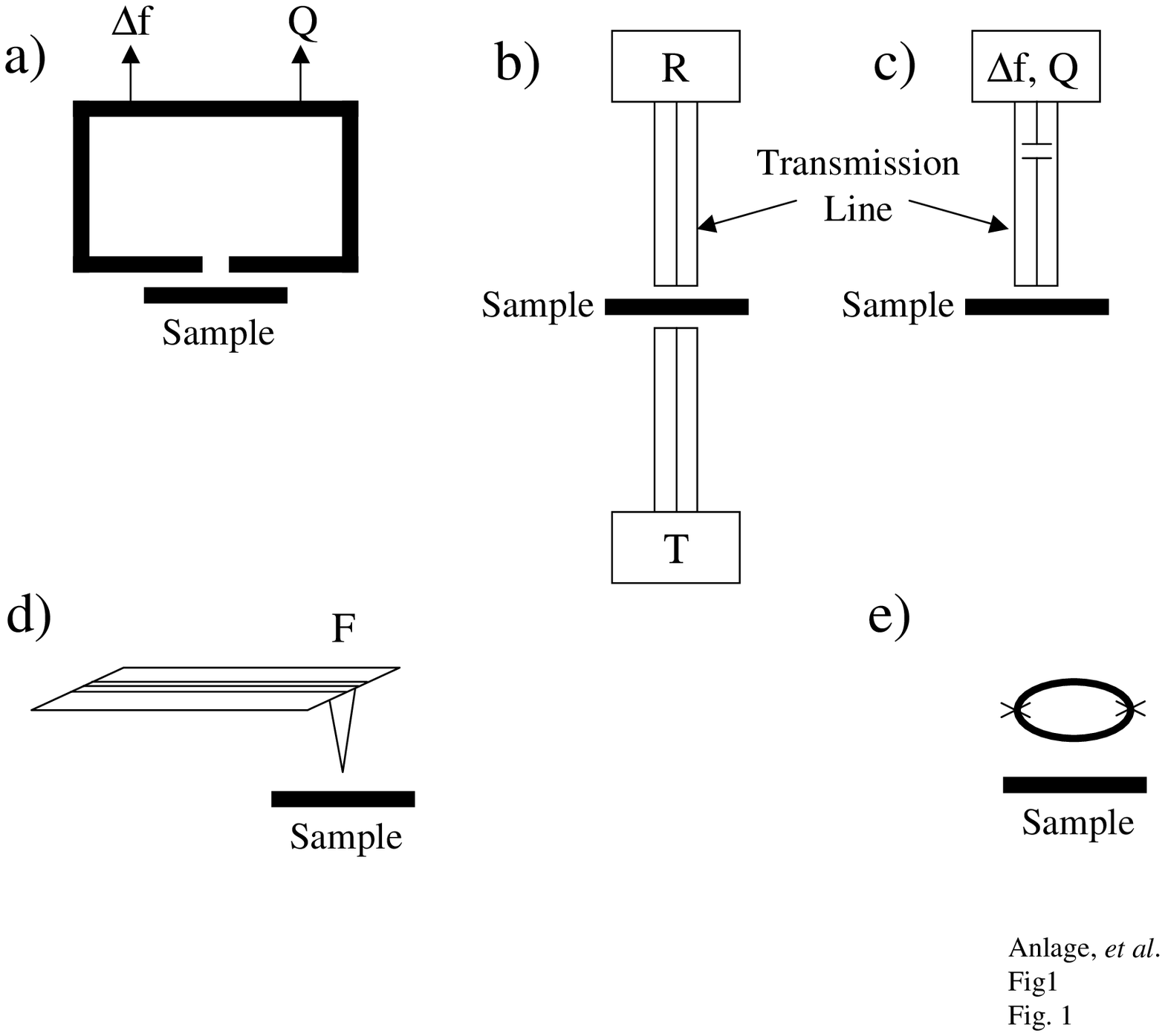,width=10cm,clip=,bbllx=50pt,bblly=347pt,
bburx=560pt,bbury=727pt}
\end{center}
\caption{Schematic of the five main classes of near-field microwave
microscopes; a) shows a microwave resonator with a small hole in one wall.
The frequency shift $\Delta $f and quality factor Q are monitored; b) shows
the transmission line reflection (R) and transmission (T) method; c) shows
the scanned transmission-line resonator technique in which $\Delta $f and Q
are monitored; d) shows the cantilevered sharp-tip method where a force (F)
or other quantities are monitored; e) shows the scanning SQUID method.}
\label{Fig1}
\end{figure}

To codify the great variety of work on near-field microwave microscopy of
materials, we present a simple picture of the five basic types of microwave
microscopy in Fig. \ref{Fig1}. Although these classifications are somewhat
gross, they allow us to cleanly distinguish the main efforts in the field.

Fig. \ref{Fig1}(a) illustrates a traditional microwave cavity resonator with
a small hole in one wall. The sample is placed in close proximity to the
wall; a small region of the sample, defined by the hole diameter, perturbs
the resonant frequency (causing a frequency shift $\Delta $f) and quality
factor (Q) of the resonator. Because the hole is so small (0.5 mm diameter
for Frait \cite{Frait}) and the sample makes such a small perturbation to
the cavity, one must examine a highly lossy property of the sample, such as
ferromagnetic resonance. Ferromagnetic resonance (FMR) gives information
about the local internal fields and magnetization of the sample. Frait \cite
{Frait}, Soohoo \cite{Soohoo} and Bhagat \cite{Lofland} have successfully
used this microscope for FMR imaging, while Ikeya \cite{Ikeya} has used it
for electron spin resonance imaging.

An important variation on this theme was developed by Ash and Nichols \cite
{Ash}. They used an open hemispherical resonator with a flat-plate
reflector. The plate had a small hole in it and the sample was positioned
close to this hole, but outside the resonator. They did not study FMR, so in
order to recover a perturbation signal from the resonator, the sample
separation from the hole was modulated at a fixed frequency. They were able
to phase-sensitively recover the reflected signal from the resonator to
construct a qualitative image of the sample as it was scanned under the
hole. These cavity methods all make use of an evanescent mode to couple the
resonant cavity to a local part of the sample. In that sense these methods
resemble the tapered optical fiber with aperture method of near-field
scanning optical microscopy (NSOM) \cite{Pohl,Betzig}.

Fig. \ref{Fig1}(b) illustrates a class of non-resonant microscopes in which
the sample is placed on or near the end of a microwave transmission line.
The complex reflectivity R or transmission coefficient T is measured, and
properties of the sample are deduced. The most common technique is to
measure reflectivity from a coaxial transmission line terminated (possibly
with an air gap) by the sample \cite
{Bryant,Xu,Stuchly,Burdette,Fee,Stranick,Jiang,Asami}. Variations include
the use of waveguides \cite{Gutman,Qaddoumi}, waveguides covered by a
resonant slit \cite{Golosovsky,Bae} and microstrip lines \cite{Tabib-Azar}
in reflection. Transmission measurements also have been done in the coaxial 
\cite{Stranick,Keilmann} and waveguide \cite{Bae} geometries. These
techniques have mainly been used to map metallic conductivity or sheet
resistance, and dielectric constant. In some cases the measurements are done
quantitatively.

Fig. \ref{Fig1}(c) shows schematically one of the most sensitive forms of
near-field microwave microscopy. The idea is that the sample is put near the
open end of a transmission-line resonator, and changes in the resonant
frequency and quality factor are monitored as the sample is scanned. This
class of techniques differs from that shown in\ Fig. \ref{Fig1}(a) through
the use of a ``field enhancing'' feature at the end of the transmission
line, rather than an evanescent aperture in the resonator. The
field-enhancing feature sets the scale for the spatial resolution of this
class of microscopes. The first embodiment of this concept was conceived for
measuring the moisture content of paper \cite{Bosisio}. Other embodiments
use coaxial transmission lines with the sample in contact with the open end 
\cite
{Tanabe,Cho,Xiang,GusFirst,AnlageASC1,DaveFirst,DaveSecond,DaveDielAPL,JohanMTT,AnlageRFSC,AnlageASC2,AnlageHFPHTSC,GusDiel,AnlageISS98,GusTopo,Werner}%
, or with an air gap between the probe and the sample \cite
{Xiang,GusFirst,AnlageASC1,DaveFirst,DaveSecond,DaveDielAPL,JohanMTT,AnlageRFSC,AnlageASC2,AnlageHFPHTSC,GusDiel,AnlageISS98,GusTopo}%
. A related far-field technique has been used to image surface resistance
and nonlinearity with a scanned dielectric resonator in contact with the
sample \cite{Gallop}.

The resonant transmission-line technique has proven to be the most
quantitative of all the near-field microwave microscopy methods.
Quantitative imaging of topography, sheet resistance, dielectric constant
and loss, and other quantities has been achieved. In addition, the use of
field enhancing features has pushed the spatial resolution of resonant and
non-resonant microscopes to the sub-$\mu $m domain while maintaining the
quantitative nature of the measurements.

The last two classes of near-field microscopy techniques are less developed,
but show great promise for the future. Fig. \ref{Fig1}(d) shows a cantilever
with a sharp tip positioned over the sample. This geometry can be used to
perform atomic force microscopy to determine the topography of the sample,
in addition to microwave microscopy. There are three sub-classes of scanning
probe measurements of materials properties at microwave frequencies. The
first is to perform localized electron-spin resonance measurements \cite
{Manassen}. The tip is used for STM, and a magnetic field is applied to the
sample. It is found that a local signal at the Larmor precession frequency
can be extracted from the tip and converted into an image of ESR response of
the sample. A second embodiment is to create a magnetic field gradient on
the sample (e.g., with a small magnetic particle on the tip) while immersing
it in an rf magnetic field. The sample will locally satisfy the magnetic
resonance condition and exert a force on the cantilever \cite{Zhang,Rugar}.
A third scanning probe method is to simply use a sharp metallized tip to
perform ``apertureless'' near-field microscopy. The sharp tip in close
proximity to a metallic sample will locally enhance radiation introduced by
a focused far-field beam. If an additional phenomenon can take place in this
localized region due to the enhanced field strength, and it can be detected,
one has a local microscopic probe of the physics associated with that
phenomenon. One important example of this kind of microscopy is apertureless
infrared microscopy, where the reflected signal from the region of the probe
tip is measured while the tip is periodically dithered up and down \cite
{Knoll}.

Fig. \ref{Fig1}(e) shows the scanning superconducting quantum interference
device (SQUID) method of microwave microscopy. The SQUID will generate
circulating rf currents when a dc bias is placed across the loop. The
frequency of these currents is directly proportional to the applied bias
voltage. The currents generate rf magnetic fields which then impinge on the
sample. The sample will generate its own response currents which in turn
modify the inductance of the SQUID loop. By monitoring the magnetic-field
feedback signal required to keep the SQUID in a constant flux state, one can
map the electromagnetic response of the sample \cite{Black}. This method has
the advantage of being very broadband (in principle from rf up to the gap
frequency of the superconductor used to make the SQUID, $\sim $100 GHz or
more) and quantitative.

\subsection{Detailed Description of the Maryland Microscope}

\begin{figure}[h]
%\vspace{5cm}   % amount of vertical space needed
\par
\begin{center}
\leavevmode
\epsfig{file=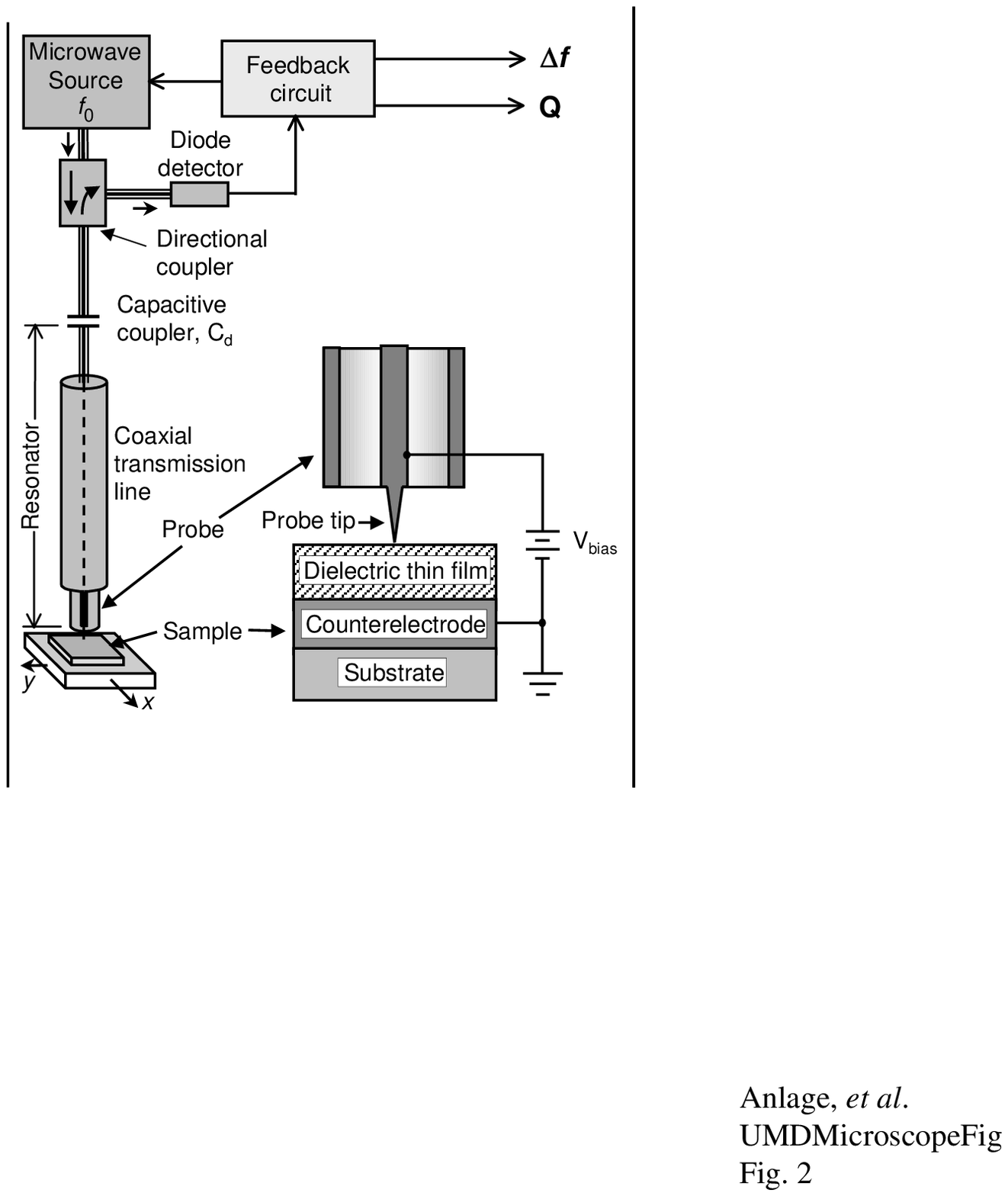,width=7cm,clip=,bbllx=164pt,bblly=378pt,
bburx=399pt,bbury=650pt}
\end{center}
\caption{A schematic diagram of the scanned resonator near-field microwave
microscope developed at the University of Maryland. The microwave source is
kept on a microscope resonance by means of a feedback loop which supplies
the frequency-shift error signal $\Delta $f.  The quality factor, Q, of the
microscope resonator is also measured.  Shown in the inset
is an expanded view of the tip/sample interaction and a dc bias circuit used
to measure local dielectric tunability.}
\label{UMDMicroscopeFig}
\end{figure}

\begin{figure}[h]
%\vspace{5cm}   % amount of vertical space needed
\par
\begin{center}
\leavevmode
\epsfig{file=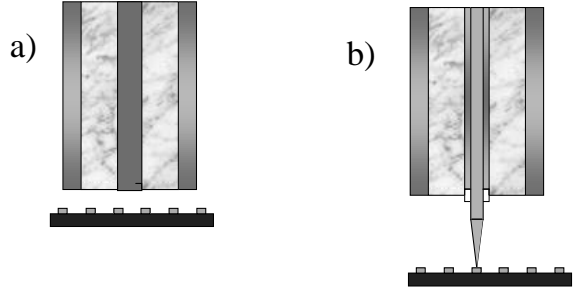,width=8cm,clip=,bbllx=92pt,bblly=269pt,
bburx=507pt,bbury=490pt}
\end{center}
\caption{Probe tips used in the scanning near-field microwave microscope. a)
shows a blunt electric-field probe tip in non-contact mode, and b) shows the
STM-tip electric-field probe in contact mode.}
\label{ProbeTipsFig}
\end{figure}

\begin{figure}[h]
%\vspace{5cm}   % amount of vertical space needed
\par
\begin{center}
\leavevmode
\epsfig{file=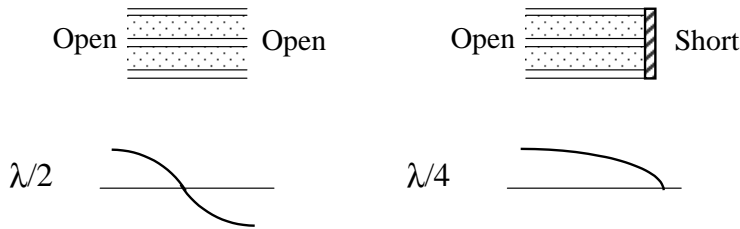,width=10cm,clip=,bbllx=75pt,bblly=492pt,
bburx=531pt,bbury=644pt}
\end{center}
\caption{Microwave microscope in the two extremes of no sample (a) and with
a metallic sample shorted to the probe (b). In (a) the microscope is a 
half-wave resonator, while in (b) it is a quarter-wave resonator.}
\label{OpenShortFig}
\end{figure}

Our group has developed a novel form of near-field scanning microwave
microscopy based on the scanned resonator technique (Fig. \ref{Fig1}(c)). As
shown in Fig. \ref{UMDMicroscopeFig}, our microscope consists of a resonant
coaxial cable which is weakly coupled to a microwave generator on one end
through a decoupling capacitor C$_d$, and coupled to a sample through an
open-ended coaxial probe on the other end. In the absence of a sample, the
microscope is a half-wave resonator (Fig. \ref{OpenShortFig}). As a metallic
sample approaches the open end of the coaxial probe, the boundary condition
there goes progressively from open circuit to short circuit (e.g., for the
blunt probe shown in Fig. \ref{ProbeTipsFig}(a)). In the limit of the sample
touching the coaxial probe, the microscope becomes a quarter-wave resonator
(Fig. \ref{OpenShortFig}). In the process, the resonant frequency shifts by
one half the spacing between neighboring modes, which is on the order of 100
MHz. Hence the microscope frequency shift is very sensitive to the
probe/sample separation, as well as the electrical properties of the sample
(i.e., the conductivity, dielectric constant and magnetic permeability).

Consider again the limit of the sample very far from the probe. In this case
the quality factor of the resonator, Q, is high since only internal
dissipation processes (and radiation losses) are active. As the sample
approaches the open end of the coaxial probe it must be considered part of
the resonant circuit. As such, it will add losses to the system and in
general reduce the Q of the microscope. The Q thus gives insight into the
additional loss mechanisms introduced by the sample.

As the sample is scanned beneath the probe, the probe-sample separation will
vary (depending upon the topography of the sample), causing the capacitive
coupling to the sample, C$_x$, to vary. This has the result of changing the
resonant frequency of the coaxial-cable resonator. Also, as the local
electrical properties of the sample vary, so will the resonant frequency and
quality factor, Q, of the resonant cable. The goal then is to turn the
measured quantities $\Delta $f and Q versus position over the sample into
quantitative images of sample electrical properties.

A circuit is used to force the microwave generator to follow a single
resonant mode of the cable, and a second circuit is used to measure the Q of
the circuit, both in real time (Fig. \ref{UMDMicroscopeFig}) \cite
{DaveFirst,DaveSecond}. The microwave source is frequency modulated by an
external oscillator at a rate $f_{FM}\sim $ 3 kHz. The electric field at the
probe tip is perturbed by the region of the sample beneath the probe's
center conductor. We monitor these perturbations using a diode detector
which produces a voltage signal proportional to the power reflected from the
resonator. A feedback circuit \cite{DaveSecond} (Fig.\ \ref{UMDMicroscopeFig}%
) keeps the microwave source locked to a resonant frequency of the
transmission line, and has a voltage output which is proportional to shifts
in the resonant frequency due to the sample. Hence, as the sample is scanned
below the open-ended coaxial probe, the frequency shift and Q signals are
collected by a computer. The circuit runs quickly enough to accurately
record at scan speeds of up to 25 mm/sec.

To determine the quality factor Q of the resonant circuit, a lock-in
amplifier, referenced at $2f_{FM}$, gives an output voltage $V_{2f_{FM}}$
which is related to the curvature of the reflected power vs. frequency curve
on resonance, and hence to Q. To relate $V_{2f_{FM}}$ and Q, we perform a
separate experiment, in which we vary Q using a microwave absorber at
various heights below the probe tip, and measure the absolute reflection
coefficient $\left| \rho \right| ^2$ of the resonator. If $\left| \rho
_0\right| ^2$ is the reflection coefficient at a resonant frequency $f_0$,
then the coupling coefficient between the source and the resonator is $\beta
=\left( 1-\left| \rho _0\right| \right) /\left( 1+\left| \rho _0\right|
\right) .$ The loaded quality factor of the resonator \cite{Aitken,Zaki} is Q%
$_L$ $=f_0/\Delta f$, where $\Delta f$ is the difference in frequency
between the two points where $\left| \rho \right| ^2=$ $\left( 1+\beta
^2\right) /\left( 1+\beta \right) ^2$. The unloaded quality factor, which is
the Q of the resonator without coupling to the microwave source and
detector, is then Q$_0=$ Q$_L\left( 1+\beta \right) $. We also measure $%
V_{2f_{FM}}$, and find that there is a unique functional relationship
between Q$_0$ and $V_{2f_{FM}}$; thus, we need to calibrate this
relationship only once for a given microscope resonance. In a typical scan,
we record $V_{2f_{FM}}$, and afterward convert $V_{2f_{FM}}$ to Q$_0$.

\section{Properties of the Near-Field Microwave Microscope}

\subsection{Spatial Resolution}

The spatial resolution of the microscope has been demonstrated to be the
larger of the probe-sample separation and the diameter of the inner
conductor wire in the open-ended coaxial cable \cite{AnlageASC1}. As with
other forms of near-field microscopy, the probe is placed well within one
wavelength of the sample under study. This is particularly easy to
accomplish at rf, microwave and millimeter-wave frequencies because the
wavelength ranges from meters to millimeters. However, the high spatial
resolution comes from a ``field-enhancement'' feature of the probe. It has
been found that sharp tips produce higher spatial resolution than blunt
probes. Figure \ref{ProbeTipsFig} contrasts these two types of probes in a
transmission-line microscope, like those shown in Fig. \ref{Fig1}(b) and
(c). Figure \ref{ProbeTipsFig}(a) shows a blunt electric field probe, while
Fig. \ref{ProbeTipsFig}(b) shows an STM-tipped probe. The blunt probe has a
spatial resolution given by the area of the center conductor when operated
in non-contact mode at a height smaller than the inner conductor diameter.
The STM tip has a spatial resolution in contact mode on the order of 1 $\mu $%
m, depending somewhat on the bluntness of the tip.

The ``lightning-rod effect'' is responsible for the increased spatial
resolution with a sharp-tip probe. This same principle is used in
apertureless near-field scanning optical microscopy. The basic idea is that
the electromagnetic fields in the near-field environment can be treated in
the static approximation. In electrostatic equilibrium there will be a large
electric field at sharp corners, with the local field being inversely
proportional to the radius of curvature. This enhancement means that the
response of the sample immediately below the sharp tip will dominate the
signal. Quantitative calculations of dielectric response from an STM-tipped
probe bear out this qualitative picture \cite{DaveSRSI}.

In addition to achieving high spatial resolution it also is important to
vary the spatial resolution while maintaining quantitative measurement
capability. There are times when a gross characterization or measurement of
a sample property is required, for instance on the scale of a wafer or
thin-film. There are physical property features on many length scales in
materials. Hence, a microscope must adapt itself to imaging all of these
scales. Our microscope is well adapted for this purpose and has been used
with spatial resolutions of 500 $\mu $m, 200 $\mu $m, 100 $\mu $m, 12 $\mu $%
m and 1 $\mu $m.

\subsubsection{Evanescence and Microwave Microscopy}

An evanescent wave is one which does not propagate, but is exponentially
attenuated with distance. Such behavior can be created with microwaves in a
variety of situations. The simplest is to use a single-conductor cylindrical
wave guide and to reduce its lateral dimensions such that the cutoff
frequency is made greater than the frequency of the propagating mode. The
wave is then attenuated on the length scale of the diameter of the waveguide 
\cite{RamoEv}. This is the method employed by the microscopes in Fig. \ref
{Fig1}(a). The attractive feature of evanescence is the fact that the wave
equation for the electromagnetic fields reduces to Laplace's equation in
regions small compared to the wavelength of the radiation. This allows one
to use low-frequency analysis to understand the distribution of the fields
in the evanescent near-field region.

Evanescence through an aperture brings with it several distinct
disadvantages. First, evanescent waveguides are characterized by a purely
reactive impedance \cite{RamoEv}, hence they reflect signals back to the
source very efficiently. This leads to the problem of a very poor
signal-to-noise ratio, as is encountered in traditional NSOM through tapered
optical fibers. The near-field optical community is now moving towards the
``apertureless NSOM'' method in which no evanescent waves are used. Instead
they rely on a field-concentrating feature to enhance the signal from a very
localized part of the sample. This is similar to the methods which we employ
in our microscopes.

Another disadvantage of evanescence through an apeture at microwave
frequencies is the long length scale of the evanescent decay. Evanescence is
of great utility in scanning tunneling microscopy because the decay length
is so short that it permits atomic resolution imaging. However, for
radiation coming out of a hole beyond cutoff, the decay length scale is on
the order of the hole diameter. This can be made no smaller than a few
hundred microns in practical microwave microscopes without reducing the
signal-to-noise ratio to unity.

However, techniques of field enhancement can improve the spatial resolution
to much smaller values without compromising the signal-to-noise. Here it may
be that short-length-scale evanescent waves associated with the field
enhancement feature may dramatically improve the spatial resolution \cite
{IchiroPC}. These evanescent waves are commonly encountered at the boundary
between two different waveguides \cite{Booth,James}. The details of how
these evanescent waves contribute to the imaging of electromagnetic
properties on short length scales remain to be determined.

\subsection{Contrast mechanisms}

The microwave microscope generates contrast for many different
electromagnetic properties of materials. These include topography,
conductivity, dielectric constant and magnetic permeability. In addition,
other properties such as ferromagnetic and antiferromagnetic resonance
produce contrast for the microscope.

In order to quantitatively extract this information from the microscope
images, it is necessary to model the microscope and its interaction with the
sample. Here we begin with a model of our microscope, and defer the
discussion of sample-specific modeling to the next section.

\begin{figure}[h]
%\vspace{5cm}   % amount of vertical space needed
\par
\begin{center}
\leavevmode
\epsfig{file=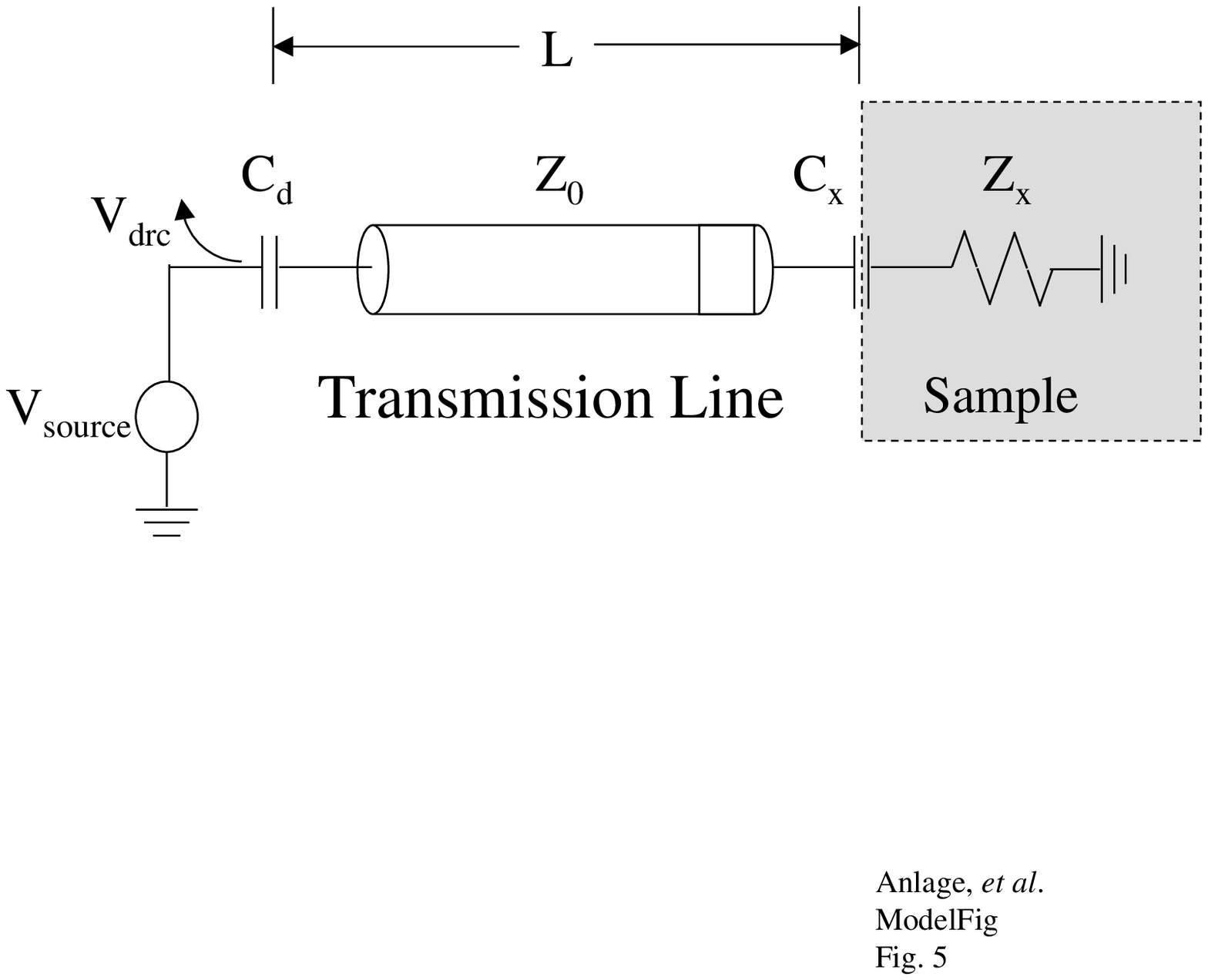,width=10cm,clip=,bbllx=60pt,bblly=322pt,
bburx=532pt,bbury=545pt}
\end{center}
\caption{Electrical schematic of a simple model used to describe the
near-field microwave microscope. The source has voltage $V_{source}$, the
coupling capacitor is $C_d$, the resonator has characteristic impedance $Z_0$
and interacts with the sample through capacitance $C_x$, and the sample has
complex impedance $Z_x$. The measured directional coupler detector output
voltage is $V_{drc}$.}
\label{ModelFig}
\end{figure}

The heart of the microscope is the microwave resonator (Fig. \ref{ModelFig}%
). It is coupled to the generator by the capacitor C$_d$ and to the sample
through the capacitor C$_x$ (in the case of an electric-field probe). The
sample can be modeled as an impedance Z$_x$ which includes resistive,
capacitive and inductive loads which it presents to the microscope. Also
included in the model are the directional coupler and detector which are
used to measure the reflected signal from the resonator.

One can calculate the signal measured in the detector V$_{drc}$ in terms of
the source driving voltage, V$_{source}$. In the simplest case, one finds
the following expression for the reflectivity of the resonator:

\begin{equation}
V_{drc}=\frac{Z_3-Z_0}{Z_3-Z_0}P(1-P)V_{source},  \label{Vdrc Eq}
\end{equation}

where Z$_3=Z_d+Z_0\frac{(Z_x+Z_0)e^{\gamma L}+(Z_x-Z_0)e^{-\gamma L}}{%
(Z_x+Z_0)e^{\gamma L}-(Z_x-Z_0)e^{-\gamma L}},$ Z$_0$ is the characteristic
impedance of the transmission line, P is the directional coupler voltage
fraction, Z$_d$ = 1/$i\omega C_d$, $\gamma $ is the complex propagation
constant of the transmission line of length L, and Z$_x$ is the impedance of
the capacitor C$_x$ in series with the sample.

The presence of the sample modifies the resonance of the microscope in
several ways. First, as the probe-sample distance varies, so does the
capacitance C$_x$. Consider a metallic sample below a blunt probe (e.g.,
Fig. \ref{ProbeTipsFig}(a)). If the sample is much closer than the probe
diameter, there is a parallel-plate capacitance between the probe and the
sample. As the sample approaches the probe, this capacitance will increase,
and the resonant frequency will decrease. This is because the increased
capacitance at the end of the transmission line will effectively lengthen
the resonator. Ultimately as the sample comes into contact with the probe, a
short-circuit condition is achieved and the microscope becomes a
quarter-wave resonator. Hence, the greatest frequency shift expected is that
from converting a half-wave to a quarter-wave resonator, which is half the
distance between neighboring modes in either the open circuit or
short-circuit microscope.

\subsection{Frequency Coverage}

Because the microscope is a transmission-line resonator with many closely
spaced modes, and because each mode can be used for imaging, the microscope
is a tremendously broadband instrument. In a typical design, the
transmission-line resonator has a length of approximately 1 m, giving a
fundamental mode at about 125 MHz, and harmonic modes at integer multiples
of this frequency. Hence, imaging can be done up to frequencies where
higher-order modes begin to propagate in the coaxial cable, which can be
over 100 GHz for small diameter cable. The only other limitation is the
microwave electronics required to operate the microscope, such as the
source, directional coupler, detector and connectors. However, using modern
sources, electronics and the 1-mm standard connector, a 100+ GHz imaging
bandwidth can be achieved.

\subsection{Temperature range}

Our microscopes work well at room temperature. However, we also have
constructed a cryogenic version of the microwave microscope. In this case
only the probe and part of the resonator is actually held at cryogenic
temperatures, along with the sample. Many images have been acquired at
liquid-nitrogen temperature (77 K) in both materials and device-imaging
modes \cite{AnlageASC2,AshfaqAPL,AshfaqASC}. We see no reason why this
cannot be extended to pumped helium temperatures (1.2 K). At the other
extreme, high-temperature coaxial cables exist which can be used for imaging
up to 1000${{}^{\circ }}$C \cite{Gershon}.

\subsection{Other Independent Parameters}

One can imagine introducing other quantities into the sample during
microwave microscopy. For instance, light illumination is an obvious degree
of freedom. We have found that measurements of tunability are very
interesting with the scanning near-field microwave microscope. In this case
the use of external electric and magnetic fields is most important.

\subsubsection{Electric field}

We have developed a technique to introduce a local and tunable electric
field in our microwave microscope when imaging in contact mode (Fig. \ref
{UMDMicroscopeFig}). A dc electric bias voltage can be applied to the center
conductor of the microwave probe through a bias tee. This will create an
electric field between the probe tip and the surrounding grounded surfaces.
By introducing a suitable ground plane, controlled electric fields can be
applied to the sample being imaged at microwave frequencies \cite
{DaveDielAPL}.

\subsubsection{Magnetic field}

We also can introduce a static magnetic field on samples being imaged by the
microwave microscope using an electromagnet. The field can be modulated for
imaging magnetic resonant phenomena in the sample.

\section{Quantitative Imaging with the Near-Field Microwave Microscope}

Here we present some of the results on quantitative imaging with our
microscope. Very simply, one expects lossy properties of the sample shown in
the model of Fig. \ref{ModelFig} to affect the Q of the microscope, while
reactive properties and topography will primarily affect the resonant
frequency.

\subsection{Topography}

Images obtained from resonant near-field scanning microwave microscopes are
the result of two distinct contrast mechanisms: shifts in the resonant
frequency due to electrical coupling between the probe and the sample, and
changes in the quality factor Q of the resonator due to losses in the sample 
\cite{DaveFirst,DaveSecond}. Such images will inevitably contain intrinsic
information about a sample (such as dielectric constant or surface
resistance) as well as extrinsic information (such as surface topography).
To facilitate quantitative imaging of intrinsic material properties, it is
essential to be able to accurately account for the effects of finite
probe-sample separation and topography.

The interaction between the probe and a metallic sample can be represented
by a capacitance C$_x$ between the sample and the inner conductor of the
probe, analogous to the mechanism in scanning capacitance microscopy \cite
{GusFirst,GusTopo}. For a highly conducting sample and a small probe-sample
capacitance C$_x$, i.e., C$_x$ $\ll $ L/cZ$_0$, the reflected voltage V$%
_{drc}$ at the output of the directional coupler can be written as above in
Eq. (\ref{Vdrc Eq}) with $Z_3=Z_d+\frac{Z_0\left[ \cosh \left( \gamma
L\right) +i\omega C_xZ_0\sinh \left( \gamma L\right) \right] }{\sinh \left(
\gamma L\right) +i\omega C_xZ_0\cosh \left( \gamma L\right) },$ where c is
the speed of light, and $\omega $ is the angular frequency of the source. A
plot of Eq. (\ref{Vdrc Eq}) versus frequency would show a series of dips
corresponding to enhanced absorption at the resonant frequencies \cite
{GusFirst}. For C$_x$=0 and C$_d$=0, the resonant frequencies are f$_n$=nc/(2%
$\sqrt{\varepsilon _r}$L), for n=1,2, ..., where $\varepsilon _r$ is the
dielectric constant of the transmission line; for our system a resonance
occurs every 125 MHz. For small C$_x$, the n-th resonant frequency changes
by:

$\Delta f_n\approx -f_n\frac{cC_xZ_0}L$.

Since C$_x$ depends on the distance between the inner conductor and the
sample, this equation implies that $\Delta $f can be used to determine the
topography of the sample. When the sample is very close to the probe, one
expects that C$_x$ $\approx $\ A$\varepsilon _0$/h, so that $\Delta f\sim
-f_nc\varepsilon _0AZ_0/hL$, where $A$ is the area of the center conductor
of the probe. On the other hand, when the sample is far from the probe, one
must resort to numerical simulation to find C$_x$(h). To find C$_x$(h) for
the coaxial probe geometry, we solved Poisson's equation using a finite
difference method on a 200 by 150 cell grid. The overall experimental
behavior is qualitatively well described by the numerical simulation,
including the weak frequency shifts observed at large separation. For
convenience, we parameterize our measured calibration curves with an
empirical function to easily transform measured frequency shifts to absolute
heights.

The topographic imaging capabilities of the system were demonstrated by
imaging an uneven, metallic sample: a U.S. quarter-dollar coin. First, the
frequency shift was recorded as a function of position over the entire
sample. We next measured the frequency shift versus height at a fixed
position above a flat part of the sample - see Fig. 2 of reference \cite
{GusTopo} - and determined the transfer function h($\Delta $f). Using h($%
\Delta $f), we then transformed the frequency-shift image into a topographic
surface plot - see Fig. 3(b) of reference \cite{GusTopo}.

As a result of this work, we can quantitatively account for the contribution
of topography to frequency-shift images. Our technique allows for a height
resolution of 55 nm for a 30-$\mu $m probe-sample separation and about 40 $%
\mu $m at a separation of 1.75 mm \cite{GusTopo}. The technique is simple
and should be readily extendible to non-metallic samples, smaller probes and
closer or farther separations.

\subsection{Sheet resistance}

Based on simple reasoning, we expect that the losses of the sample will
affect the quality factor Q of the microscope. A simple situation to study
is that of a resistive thin-film on a dielectric substrate with a thickness
much less than the skin depth. In this case one can show that the sample
presents a sheet resistance R$_x$ = $\rho $/t to the microscope, where $\rho 
$ is the thin-film resistivity and t is the local thickness of the film. To
determine the relationship between Q$_0$ and sample sheet resistance ($R_X$%
), we used a variable-thickness aluminum thin-film on a glass substrate \cite
{DaveFirst}. The cross-section of the thin-film is wedge-shaped, implying a
spatially-varying sheet resistance. Using a probe with a 500-$\mu $%
m-diameter center conductor in non-contact mode, and selecting a resonance
of the microscope with a frequency of 7.5 GHz, we acquired frequency-shift
and Q$_0$ data. We then cut the sample into narrow strips to take two-point
resistance measurements and determine the local sheet resistance. We find
that Q$_0$ reaches a maximum as $R_X\rightarrow 0$; as $R_X$ increases, Q$_0$
drops due to loss from currents induced in the sample, reaching a minimum
around $R_X=660\ \Omega /\Box $ for a height of 50 $\mu $m. Similarly, as $%
R_X\rightarrow \infty $, Q$_0$ increases due to diminishing currents in the
sample. This means that $R_X$ is a double-valued function of Q$_0$. This
presents a problem for converting the measured Q$_0$ to $R_X$. However, $R_X$
is a single-valued function of the frequency shift \cite{DaveFirst},
allowing one to use the frequency-shift data to determine which branch of
the $R_X($Q$)$ curve should be used.

To explore the capabilities of our system, we scanned a thin-film of YBa$_2$%
Cu$_3$O$_{7-\delta }$ (YBCO) on a 5-cm-diameter sapphire substrate at room
temperature \cite{DaveSecond}. The film was deposited using pulsed laser
deposition with the sample temperature controlled by radiant heating. The
sample was rotated about its center during deposition, with the $\sim $%
3-cm-diameter plume held at a position halfway between the center and the
edge. The thickness of the YBCO thin-film varied from about 100 nm at the
edge to 200 nm near the center.

\begin{figure}[h]
%\vspace{5cm}   % amount of vertical space needed
\par
\begin{center}
\leavevmode
\epsfig{file=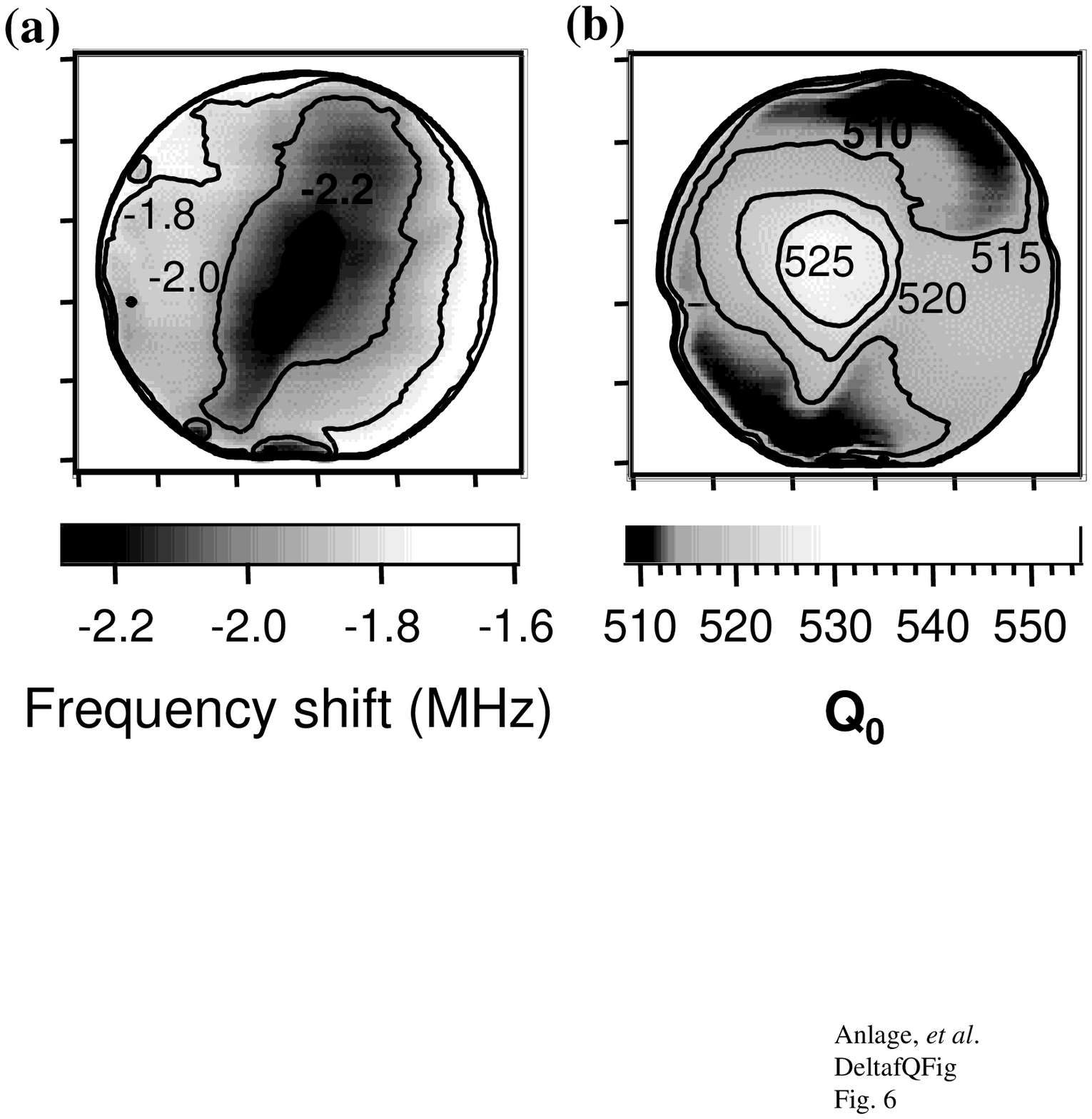,width=10cm,clip=,bbllx=65pt,bblly=286pt,
bburx=560pt,bbury=630pt}
\end{center}
\caption{Images of (a) frequency shift and (b) quality factor Q of a 
2-inch-diameter variable thickness YBa$_2$Cu$_3$O$_7$ thin film on sapphire. The
frequency-shift image is relative to the probe far away ($>$ 1 mm) from the sample. A
probe with a 500-$\mu$m-diameter center conductor was used at a height of
approximately 50 $\mu$m, at a frequency of 7.5 GHz.}
\label{DeltafQFig}
\end{figure}

Figure\ \ref{DeltafQFig} shows two microwave images of the YBCO sample. The
frequency shift [Fig.\ \ref{DeltafQFig}(a)] and Q$_0$ [Fig.\ \ref{DeltafQFig}%
(b)] were acquired simultaneously, using a probe with a 500-$\mu $m-diameter
center conductor at a height of approximately 50 $\mu $m above the sample.
The scan took approximately 10 minutes to complete, with raster lines 0.5 mm
apart. The frequency shifts in Fig. \ref{DeltafQFig}(a) are relative to the
resonant frequency of 7.5 GHz when the probe was far away ($>$1 mm) from the
sample; the resonant frequency shifted downward by more than 2.2 MHz when
the probe was above the center of the sample. Noting that the resonant
frequency drops monotonically between the edge and the center of the film,
and that the resonant frequency is a monotonically increasing function of
sheet resistance \cite{DaveFirst}, we conclude that the sheet resistance
decreases monotonically between the edge and the center.

\begin{figure}[h]
%\vspace{5cm}   % amount of vertical space needed
\par
\begin{center}
\leavevmode
\epsfig{file=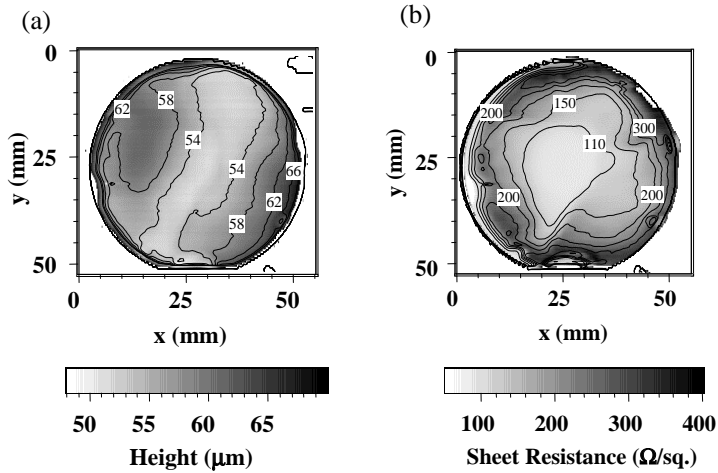,width=10cm,clip=,bbllx=78pt,bblly=111pt,
bburx=748pt,bbury=545pt}
\end{center}
\caption{Images of (a) topography (in $\mu$m below the probe tip) and (b)
sheet resistance of a 2-inch-diameter variable-thickness YBa$_2$Cu$_3$O$_7$
thin film on sapphire. }
\label{RxTopoFig}
\end{figure}

The frequency shift and Q$_0$ images [Fig.\ \ref{DeltafQFig}(a) and (b)]
differ slightly in the shape of the contour lines. This is due to the 300-$%
\mu $m-thick substrate being warped, causing a variation of a few microns in
the probe-sample separation during the scan. In practice, the frequency
shift ($\Delta $f) is dominated by topography, but also has contributions
from R$_x$. The change in Q is dominated by sheet resistance, but also has
contributions from the topography. To deal with this, we have devised a way
to deconvolve these influences and produce images of sheet resistance which
are not contaminated by topographic features, and topographic images which
are not contaminated by sheet resistance variations. The results are shown
in Fig. \ref{RxTopoFig}. Figure\ \ref{RxTopoFig}(b) confirms that the film
does, indeed, have a lower sheet resistance near the center, as was intended
when the film was deposited. We note that the sheet resistance does not have
a simple radial dependence, due to either non-stoichiometry or defects in
the film. Figure\ \ref{RxTopoFig}(a) shows that the wafer is, indeed,
warped, with a higher ridge running from upper right to lower left.

After scanning the YBCO film, we patterned it and made four-point dc
resistance measurements throughout the wafer. The dc sheet resistance had a
spatial dependence identical to the microwave data in Fig.\ \ref{RxTopoFig}%
(b) \cite{AnlageASC2}. However, the absolute values were approximately twice
as large as the microwave results, most likely due to degradation of the
film during patterning. We also found from surface profilometer traces that
the topography image of Fig. \ref{RxTopoFig}(a) is quantitatively correct.

We have estimated the sheet resistance sensitivity for the microscope as $%
\Delta R_X/R_X$ = $6.4\times 10^{-3}$, for $R_X=100\ \Omega /\Box $ using a
probe with a 500-$\mu $m-diameter center conductor at a height of 50 $\mu $m
and a frequency of 7.5 GHz \cite{DaveSecond}. The sensitivity scales with
the capacitance between the probe center conductor and the sample ($C_X$);
increasing the diameter of the probe center conductor and/or decreasing the
probe-sample separation would improve the sensitivity.

\subsection{Dielectric Constant}

The ability to image variations in relative permittivity or dielectric
constant $\varepsilon _r$ is useful for both fundamental and applied
reasons. For example, the dielectric properties of thin-film ferroelectric
materials are of interest in studying the finite-size effect on the
ferroelectric phase transition. In thin-film microelectronics, testing for
variations in dielectric constant can be used for quality control or to
develop better growth techniques. Also, knowledge of the dielectric constant
at microwave frequencies is of great importance for the design of broadband
circuits and future generations of high-speed microprocessors.

\subsubsection{Linear dielectric response - Non-Contact Mode}

We began our studies of dielectric materials with measurements in
non-contact mode with a blunt probe (Fig. \ref{ProbeTipsFig}(a)) \cite
{GusDiel}. To calibrate the system for dielectric measurements, we
constructed a test sample by placing six pieces of different dielectric
material into the bottom of a square plastic mold and pouring epoxy into the
mold. In addition, silicone adhesive was used to hold down each piece. After
the epoxy cured, the test sample was removed from the mold, polished and
positioned on the XY table. The materials embedded in the epoxy were
silicon, glass microscope slide, SrTiO$_3$, Teflon, sapphire and LaAlO$_3$.
All six pieces were approximately 500-$\mu $m thick and about 6 mm x 8 mm in
size. The overall thickness of the test sample was 6 mm.

We measured the frequency shift $\Delta $f versus height h above the six
pieces, which have dielectric constants ranging from 2.1 to about 230. We
also tested the epoxy which has an unknown dielectric constant. Each piece,
as well as the probe, was flat and smooth on the scale of 5 $\mu $m, as
judged by an optical microscope. For these measurements, we used a probe
with a 480-$\mu $m center-conductor diameter and a source frequency of 9.08
GHz. For each scan, the probe was first brought in contact with a dielectric
and the frequency shift $\Delta $f was recorded as the height was
systematically increased. Samples with the largest dielectric constant
produced the largest frequency shift, as expected \cite{GusDiel}. The
largest shift we observed was $-$26.2 MHz, when the probe was in contact
with a SrTiO$_3$ sample with $\varepsilon _r$ $\sim $ 230 \cite{GusDiel}.
The smallest shift we found was -1.2 MHz when the probe was in contact with
a Teflon sample with $\varepsilon _r$ $\sim $ 2.1 \cite{GusDiel}. The
frequency shift is essentially zero above 1 mm and saturates when the
probe-sample distance is smaller than a few microns.

We used the above information to construct an empirical calibration curve
that directly relates the frequency shift to the dielectric constant $%
\varepsilon _r$. In order to construct the calibration curve we took the
difference between the frequency shift at two different heights, h$_1$ and h$%
_2$, i.e., f$_d$=$\Delta $f(h$_2$)-$\Delta $f(h$_1$), where h$_2$ is far
away (h$_2$ $>$ 1000 $\mu $m). By taking the difference, we
eliminated the effect of drift in the microwave source frequency. Proceeding
this way for the test samples, we constructed two calibration curves of f$_d$
versus $\varepsilon _r$ - see Fig. 2 of reference \cite{GusDiel} - one curve
for h$_1$=10 $\mu $m and h$_2$=1.1 mm, and the other for h$_1$=100 $\mu $m
and h$_2$=1.1 mm. We then parameterized each calibration curve with an
empirical function, allowing us to easily transform any measured frequency
shift to a dielectric constant. From these curves we found that we can
enhance the sensitivity to the dielectric constant considerably by using a
small probe height. On the other hand, at closer probe-sample separations
the influence of topographic features will be enhanced. Hence, it is very
important to either control the height of the probe or to de-convolve the
topography from the resulting frequency shift and Q images taken in
non-contact mode.

\subsubsection{Linear dielectric response - Contact Mode}

To avoid the issue of topographic features corrupting the dielectric
imaging, we have developed a contact-mode version of dielectric imaging \cite
{DaveDielAPL}. We use the near-field scanning microwave microscope with a
sharp protruding center conductor. The probe tip, which has a radius of
curvature $\sim 1\ \mu $m - see inset to Fig.\ \ref{UMDMicroscopeFig} and
Fig. \ref{ProbeTipsFig}(b) - is held fixed, while the sample is supported by
a spring-loaded cantilever applying a controlled normal force of about 50 $%
\mu $N between the probe tip and the sample \cite{DaveDielAPL}. Due to the
concentration of the microwave fields at the tip, the boundary condition of
the resonator, and hence, the resonant frequency $f_0$ and quality factor Q
are perturbed depending on the dielectric properties of the region of the
sample immediately beneath the probe tip. We have shown that the spatial
resolution of the microscope in this mode of operation is about 1 $\mu $m 
\cite{AnlageASC1}.

To observe the microscope's response to sample dielectric permittivity $%
\epsilon _r$, we monitored the frequency-shift signal while scanning samples
with known $\epsilon _r$. With well-characterized 500-$\mu $m-thick bulk
dielectrics, we observed that the microscope frequency shift monotonically
increases in the negative direction with increasing sample permittivity, as
noted above \cite{GusDiel}. We observed similar behavior with thin-film
dielectric samples.

For comparison, we did a finite-element calculation of the rf electric field
near the probe tip. Because the probe-tip radius is much less than the
wavelength $(\lambda \sim 4$ cm at 7 GHz), a static calculation of the
electric field is sufficient \cite{RamoEv}. Cylindrical symmetry further
simplifies the problem to two dimensions. We represent the probe tip as a
cone with a blunt end, held at a potential of $\phi =1$ V. Using relaxation
methods we solved Poisson's equation for the potential, $\nabla ^2\phi =0$,
on a rectangular grid representing the region around the probe tip. The two
variables we used to represent the properties of the probe were the aspect
ratio of the probe tip ($\alpha \equiv dz/dr$) and the radius $r_0$ of the
blunt end.

Since the sample represents a small perturbation to the resonator, we can
use perturbation theory \cite{DaveDielAPL} to find the change in the
resonant frequency: 
\begin{equation}
\frac{\Delta f}f\approx \frac{\epsilon _0}{4W}\int_{V_S}(\epsilon
_{r2}-\epsilon _{r1}){\bf E}_1\cdot {\bf E}_2dV,  \label{delta_f_equation}
\end{equation}
where ${\bf E}$$_1$ and ${\bf E}$$_2$, and $\epsilon _{r1}$ and $\epsilon
_{r2}$ are the unperturbed and perturbed electric fields, and relative
permittivities of two samples, respectively; $W$ is the energy stored in the
resonator, and the integral is over the volume of the sample. We compute an
approximate $W$ using the fact that the loaded quality factor \cite
{DaveSecond} of the resonator is $Q_L=\omega _0W/P_{loss}$, where $\omega _0$
is the resonant frequency, and $P_{loss}$ is the power loss in the
resonator. Using four bulk samples with known relative dielectric
permittivities between 2.1 and 305, and fixing $r_0=(0.6\ \mu $m$)/\alpha $,
we used $\alpha $ as a fitting parameter to obtain agreement between the
model results from Eq.\ (\ref{delta_f_equation}) and our data at 7.2 GHz; we
found agreement to within 10 \% for several different probe tips, with $%
1.0<\alpha <1.7$.

To extend this calibration model to thin films, we extend the finite-element
calculation to include a thin-film on top of the dielectric sample
substrate. Once a probe's $\alpha $ parameter is determined using the bulk
calibration described above, we use the thin-film model combined with Eq.\ (%
\ref{delta_f_equation}), integrating over the volume of the thin-film, to
obtain a functional relationship between $\Delta f$ and the dielectric
permittivity of the thin-film. Using the thin-film model, we found that for
high-permittivity ($\epsilon _r>50$) thin films, the microwave microscope is
primarily sensitive to the in-plane component of the permittivity tensor.

\begin{figure}[h]
%\vspace{5cm}   % amount of vertical space needed
\par
\begin{center}
\leavevmode
\epsfig{file=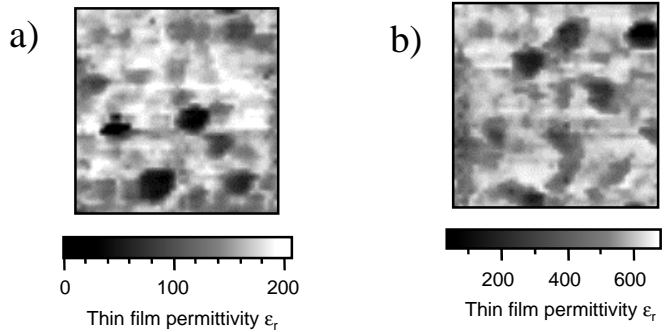,width=9cm,clip=,bbllx=83pt,bblly=302pt,
bburx=500pt,bbury=518pt}
\end{center}
\caption{a) Dielectric constant of thin-film SrTiO$_3$ over a 20-$\mu$m by
20-$\mu$m area taken at 7.2 GHz. The substrate is LaAlO$_3$, and the 
440-nm-thick film was made by pulsed laser deposition. b) Dielectric constant of
thin-film Ba$_{0.60}$Sr$_{0.40}$TiO$_3$ over a 20-$\mu$m by 20-$\mu$m area
taken at 7.2 GHz. The substrate is LaAlO$_3$ and the 275-nm-thick film was
made by pulsed laser deposition}
\label{STOEpsFig}
\end{figure}

Figure \ref{STOEpsFig}(a) shows a quantitative permittivity $\epsilon _r$
image of a sample consisting of a 440-nm SrTiO$_3$ (STO) thin-film on a 500-$%
\mu $m LaAlO$_3$ (LAO) substrate. The film was made by pulsed laser
deposition at 700$^{\circ }$ C, in 200 mTorr of O$_2$. The film is
paraelectric at room temperature. In the microwave (7.2 GHz) permittivity
image [Fig. \ref{STOEpsFig}(a)], the film shows a dielectric constant on the
order of 180 over most of its area (20 $\mu $m by 20 $\mu $m here). Several
low-permittivity defects on the film also are visible. From atomic force
microscopy measurements we find that these features are second-phase laser
particles deposited on top of the STO film. Their greater thickness is part
of the reason the relative permittivity is lower in the image.

Figure \ref{STOEpsFig}(b) shows a quantitative permittivity $\epsilon _r$
image of a sample consisting of a 275-nm Ba$_{0.60}$Sr$_{0.40}$TiO$_3$ (BST)
thin-film on a 500-$\mu $m LAO substrate. The film also was made by pulsed
laser deposition at 700$^{\circ }$ C, in 200 mTorr of O$_2$. The film is
paraelectric at room temperature. It shows a larger average dielectric
constant than STO, but also contains laser particles. The fine structure in
Figs. \ref{STOEpsFig} may be indicative of varying dielectric properties
from grain to grain in the films.

\subsubsection{Nonlinear dielectric response in contact mode}

In order to measure the local dielectric tunability of thin films, we apply
a dc electric field to the sample by voltage biasing ($V_{bias}$) the probe
tip - see Fig.\ \ref{UMDMicroscopeFig}. A grounded metallic counterelectrode
layer immediately beneath the dielectric thin-film acts as a ground plane.
To prevent the counterelectrode from dominating the microwave measurement,
thus minimizing its effect on the microwave fields - we ignored the
counterelectrode in our static field model - the sheet resistance of the
counterelectrode should be as high as possible. In our case, we use low
carrier density La$_{0.95}$Sr$_{0.05}$CoO$_3$ with a thickness of 100 nm,
giving a sheet resistance $\sim 400\ \Omega \slash \Box $. We have confirmed
by experiment and model calculation \cite{DaveFirst,DaveDielAPL} that the
contribution of the counterelectrode to the frequency shift is small ($%
\Delta f<30$ kHz) relative to the contribution from a dielectric thin-film
with thickness $>$ 100 nm ($\Delta f>200$ kHz).

\begin{figure}[h]
%\vspace{5cm}   % amount of vertical space needed
\par
\begin{center}
\leavevmode
\epsfig{file=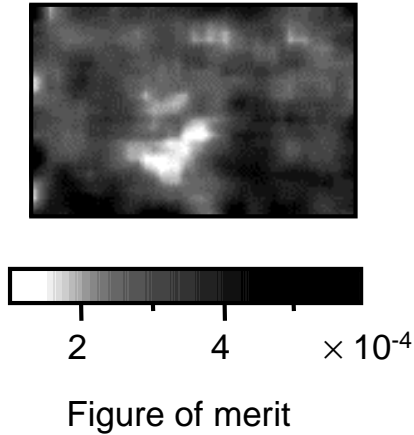,width=6.0cm,clip=,bbllx=220pt,bblly=437pt,
bburx=376pt,bbury=602pt}
\end{center}
\caption{Figure of merit of a 400-nm-thick Ba$_{0.60}$Sr$_{0.40}$TiO$_3$
film on LaAlO$_3$ made by pulsed laser deposition. The measurement is made
at 7.2 GHz and shows a 20-$\mu$m by 13-$\mu$m area. The figure of merit is
defined as the relative frequency shift of the microscope due to dc electric
field bias multiplied by the Q signal.}
\label{FOMFig}
\end{figure}

We can examine the tunability of the dielectric properties at a fixed point
on the sample and generate a hysteresis loop \cite{DaveDielAPL,DaveSRSI}. As
expected, the permittivity goes down and the Q goes up when a voltage is
applied. To image the tunability of the sample, we change the bias voltage
applied to the probe tip and image the dielectric constant again. A
tunability figure of merit can be defined with this data. In general, one
wants a highly tunable material which is not very lossy. Hence, the figure
of merit to maximize is K = ($\Delta \varepsilon _r$/$\varepsilon _r$)/tan$%
\delta $, where $\Delta \varepsilon _r$ is the change in dielectric constant
upon tuning with some fixed electric field, and tan$\delta $ is the
dielectric loss at zero bias. In our case we can define a similar figure of
merit with the raw data as K = ($\Delta f$/$f$)$Q$, where $\Delta f$ is the
frequency shift due to tuning a 400-nm-thick Ba$_{0.60}$Sr$_{0.40}$TiO$_3$
(BST) thin-film with a bias of 2.5 V, and Q is the unbiased quality factor
of the microscope. Figure \ref{FOMFig} shows the resulting figure of merit
image over a 20-$\mu $m by 13-$\mu $m area. The main region of very low
tunability (white area) is most likely a laser-deposited particle of
non-paraelectric material. However, there are smaller regions of low
tunability which are separated by just a few microns from regions of very
good tunability. This demonstrates the power of the scanning microwave
microscope to delve into the microstructure-property relationship on the
microscopic scale.

The microwave technique we use is sensitive to both film thickness and
permittivity. As a result, the permittivity of the large defects in Fig.\ 
\ref{STOEpsFig}(a) is underestimated due to the change in film thickness at
these locations, which we have confirmed with our model. However, by
examining an AFM image, topographic features can be readily distinguished
from permittivity features. We have further shown that in most of the images
the topography is too small to account for a significant contribution to the
observed permittivity contrast. We can calculate the sensitivity of the
microwave microscope by observing the noise in the dielectric permittivity
and tunability data. For a 370-nm-thick film on a 500-$\mu $m-thick LAO
substrate, with an averaging time of 40 ms, we find that the relative
dielectric permittivity sensitivity is $\Delta \epsilon _r=2$ at $\epsilon
_r=500$, and the tunability sensitivity is $\Delta (d\epsilon _r/dV)=0.03$ V$%
^{-1}$ \cite{DaveDielAPL,DaveSRSI}

\section{Future Prospects}

Having demonstrated quantitative imaging of losses, topography and
dielectric constant on the micron length scale and below, what is next for
this impressive instrument? Ongoing research is focused on a number of
issues. The first is an effort to improve the spatial resolution of the
microscope while maintaining its quantitative capabilities. The frontier
seems to be improved field-enhancement techniques, similar to those now
being pursued in near-field scanning optical microscopy. The exploitation of
resonant electromagnetic phenomena combined with cantilever techniques is
another tool for improving spatial resolution. Another important issue is
the utilization of the microwave microscope's broad frequency bandwidth.
This can be used to examine frequency-dependent phenomena, such as
electronic and magnetic dynamics in correlated electron systems.

Superconducting materials present an interesting problem for near-field
microwave microscopes. Already progress has been made in imaging the
transition temperature in superconducting thin films \cite
{Xiang,JohanMM99,Lann1,Lann2}. Although a superconducting microscope is
probably required to image surface impedance on the micron scale, it may be
possible to learn more from conventional microscopes which study
nonlinearities \cite{AnlageHFPHTSC,AnlageISS98,Wensheng}. Our group also has
imaged electric fields in the near-field of operating superconducting
microwave devices at 77 K and above \cite
{Wensheng,AshfaqAPL,SudeepAPL,AshfaqASC}. The effort is focused on isolating
the local sources of nonlinearity under operating conditions.

Many other possibilities of electrodynamics experiments remain to be
exploited. What is clear is that a new era of local electromagnetic
experiments has begun.

\section{Conclusions}

We have given a short introduction to the new field of near-field microwave
electrodynamics measurements. Five major developments of near-field
microwave microscopy have been identified. We have presented a thorough
introduction to the form of microscopy we have developed at the University
of Maryland. Quantitative images of metallic sheet resistance, topography,
dielectric constant and dielectric tunability have been presented. Great
potential exists for new quantitative electrodynamics measurements on ever
finer length scales.

\subsection{Acknowledgements}

This work could not have been possible without the assistance of S. K.
Dutta, B. J. Feenstra, W. Hu, A. Schwartz, J. Lee and A. Thanawalla. This
work has been supported by the Maryland/NSF Materials Research Science and
Engineering Center on Oxide Thin Films DMR 9632521, as well as NSF grant \#
ECS-9632811, an NSF SBIR (DMI-9710717) subcontract from Neocera, Inc. and by
the Maryland Center for Superconductivity Research.

% References

\end{document}